\documentclass{article}

\usepackage{arxiv}

\usepackage{subcaption}
\usepackage[utf8]{inputenc} 
\usepackage[T1]{fontenc}    
\usepackage{url}            
\usepackage{booktabs}       
\usepackage{amsfonts}       
\usepackage{nicefrac}       
\usepackage{microtype}      
\usepackage{lipsum}		
\usepackage{graphicx}
\usepackage{natbib}
\usepackage{doi}
\usepackage{amssymb}
\usepackage{booktabs}
\usepackage{amsmath}
\usepackage{hyperref}       
\title{Multi-Channel Auto-Encoders and a Novel Dataset for Learning Domain Invariant Representations of Histopathology Images}
\DeclareMathOperator*{\argmin}{arg\,min}

\author{Andrew Moyes$^1$\and
    Richard Gault$^1$\and
    Kun Zhang$^2$\and
    Ji Ming$^1$\and
    Danny Crookes$^1$\and
    Jing Wang$^3$}
\date{}

\renewcommand{\shorttitle}{\textit{MCAE}}

\hypersetup{
pdftitle={Multi-Channel Auto-Encoders and a Novel Dataset for Learning Domain Invariant Representations of Histopathology Images},
pdfauthor={Andrew Moyes, Richard Gault, Kun Zhang, Ji Ming, Danny Crookes, Jing Wang},
pdfkeywords={Stain Invariance, Representation Learning, Deep Learning, Histopathology},
}

\begin{document}
\maketitle
\vspace{-1cm}
\begin{enumerate}
    \item School of Electronics, Electrical Engineering and Computer Science, Queen's University, Belfast,\\ Northern Ireland, UK. email:\{amoyes01,richard.gault, j.ming, d.crookes\}@qub.ac.uk
    \item School of Electrical Engineering, Nantong University, Nantong, China. email: zhangkun\_nt@163.com
    \item Second People's Hospital of Nantong, China. email: 654383250@qq.com
\end{enumerate}
\vspace{24pt}
\begin{abstract}
	 Domain shift is a problem commonly encountered when developing automated histopathology pipelines. The performance of machine learning models such as convolutional neural networks within automated histopathology pipelines is often diminished when applying them to novel data domains due to factors arising from differing staining and scanning protocols. The Dual-Channel Auto-Encoder (DCAE) model was previously shown to produce feature representations that are less sensitive to appearance variation introduced by different digital slide scanners. In this work, the Multi-Channel Auto-Encoder (MCAE) model is presented as an extension to DCAE which learns from more than two domains of data. Additionally, a synthetic dataset is generated using CycleGANs that contains aligned tissue images that have had their appearance synthetically modified. Experimental results show that the MCAE model produces feature representations that are less sensitive to inter-domain variations than the comparative StaNoSA method when tested on the novel synthetic data. Additionally, the MCAE and StaNoSA models are tested on a novel tissue classification task. The results of this experiment show the MCAE model out performs the StaNoSA model by 5 percentage-points in the f1-score. These results show that the MCAE model is able to generalise better to novel data and tasks than existing approaches  by actively learning normalised feature representations.
\end{abstract}


\section{Introduction}
The introduction of digital whole-slide scanners has given rise to the field of digital histopathology. Alongside digital histopathology, the volume of digital tissue images has grown enormously. In order to process this wealth of data, numerous image analysis algorithms have been produced \citep{gurcan2009histopathological}. However, many of these algorithms are dataset specific, meaning they may not perform well if they are applied to datasets that are significantly different to those for which they were created. Image preprocessing is very important in histopathology due to the inter-domain differences, such as inter-scanner differences and differences in staining protocols.

Although digital whole slide scanners have enabled the production of enormous datasets that are useful for data mining and machine learning, the differences in the underlying capture technology can result in very different colour distributions in the final images even when digitising the same tissue specimen. This can lead to a requirement for specific models to be trained for each scanner. Modern machine learning methodologies such as convolutional neural networks typically capture colour and texture information in the initial layers of the network \citep{krizhevsky2012imagenet}. Different slide scanners can introduce significant variance in colour, therefore a neural network that has been trained only on data from one scanner will likely suffer from performance degradation if applied to data captured using a different type of scanner \citep{ciompi2017importance}. Global colour transformations have been used to normalise the colour distributions of images from different scanners but these often introduce artifacts due different proportions of anatomical structures \citep{janowczyk2017stain}. This problem could be avoided by calibrating a model based upon data obtained by the same tissue specimen being digitally captured by the different scanners, however this type of data is very rare. Even when this data can be acquired, tissue registration (or image alignment) between scanners can be difficult due to differences in the underlying capture technology that can mean a simple linear translation in the $x$ and $y$ axes is not sufficient to achieve a whole slide alignment.

In this work these challenges are tackled by extending an earlier method developed by the authors called the Dual-Channel Auto-Encoder model (DCAE) \citep{moyes2018novel} to model data that, theoretically, could derive from $N$ domains through the development of the Multi-Channel Auto-Encoder (MCAE) model. The major contributions of this work are:

\begin{itemize}
    \item The development of a novel synthetic dataset using CycleGANs, which contains representations of images in new known domains. This provides a clear ground truth to evaluate computational models that need to be applied in multiple domains.
    \item The development of a novel auto-encoder model based on dual-channel auto-encoders to learn domain-invariant feature representations from $N$ domains of histopathology images. This methodology enhances the expanding area of classification modelling within digital histopathology to account for inter-scanner (or inter-laboratory) variations that might restrict the generalisation of automated digital histopathology analysis.
\end{itemize}

The rest of this paper is organised as follows: Section \ref{sec:background} provides a review of the relevant literature and an outline of the context in which this work is proposed. Section \ref{sec:cyclegan_generated_synthetic_dataset} outlines the methods used to develop the synthetic dataset, which is the first contribution of this paper. The second contribution of this work is the proposed MCAE which is outlined in Section \ref{sec:multi_channel_auto_encoder} and utilises the synthetic dataset from Section \ref{sec:cyclegan_generated_synthetic_dataset}. In Section \ref{sec:experimental_eval} the quality of the synthetic dataset is evaluated in terms of its suitability for use in further experiments. Additionally, the MCAE is evaluated and compared with the StaNoSA model \citep{janowczyk2017stain}. The paper concludes with a discussion in Section \ref{sec:conclusion}.

\section{Background}\label{sec:background}
During the staining process, the individual standards and protocols of each laboratory can lead to deviations in average colour intensity of stains and overall image appearance \citep{ciompi2017importance}. Additionally, factors such as ambient temperature, time in staining solution and improper cleaning can all affect the end result with respect to tissue staining. In a similar fashion to the inter-scanner variances, these differences in staining conditions can result in dramatically different colour distributions across domains and therefore cause problems for image processing algorithms. A common approach to normalising the stain colours of an image is to first apply stain separation. Given a tissue section stained with hematoxylin and eosin, stain separation aims to decompose the single image stain with hematoxylin and eosin into two images where each contains only a single stain. Global colour transformations can then be applied to each single-stain image separately to normalise the colours to that of another domain \citep{magee2009colour}. Some stain separation algorithms are limited by the need for expert parameter definition \citep{macenko2009method}. The more pressing problem is that there are very few datasets available to obtain such information because the process of producing a dataset where each specimen has been stained with hematoxylin and eosin as well as each individual stain is time consuming, meaning normalisation methods based on stain separation can be unreliable.

Deep learning has provided radical improvements in the achievable results of many histopathology-related tasks including cell segmentation, tumour detection, mitotis detection and colour normalisation \citep{janowczyk2016deep, sirinukunwattana2016locality, komura2018machine}. Generative adversarial networks (GANs)  \citep{goodfellow2014generative} are a type of deeply-learned model that enable the sampling of novel data points given a dataset of real images. GANs are comprised of two differentiable models - the generator and the discriminator. The discriminator learns a function that maps observed data points (e.g. images) to scalar values and updates its parameters such that the values assigned to observations sampled from the set of real images are maximised whilst the values assigned to observations drawn from the generator function are minimised. The generator function approximates the real data distribution by conditioning the observations $x$ on latent variables $z$. Producing artificial or ``fake" samples is achieved by sampling from this conditional distribution $p_{model} = p(x|z)$. The parameters of the generator are updated in the direction which maximises the score assigned to the fake samples; thus an adversarial minimax game is defined where the discriminator tries to minimise the score assigned to samples drawn from the generator and the generator is trying to maximise the score indirectly. 
Conditional generative adversarial networks (cGANs) \citep{isola2017image} are a modification of the traditional GAN formulation where the generator receives additional information about the sample to be generated. For example, given a task of generating hand-written digits from the MNIST dataset, the observations are conditioned on some representation of the class $y$ (e.g. a one-hot vector) as well as the latent variable $z$, i.e. the generator is modelling $p(x|y,z)$. A method for mapping colour appearances between domains of histopathology images called Stain Style Transfer \citep{cho2017neural} used cGANs with an additional feature-preserving loss to enable mapping of stain colour appearance without corrupting the underlying tissue structure of transferred images.

Cycle-consistent generative adversarial networks, or CycleGANs \citep{zhu2017unpaired} are a type of cGAN but rather than conditioning the generative process on a class label, it is conditioned on an image. These models are useful as they are able to learn transformations between domains of images without requiring the datasets to be aligned or corresponding in any way. CycleGANs have proven highly effective in a number of colour normalisation scenarios within the scope of histopathology images. The StainGAN model \citep{shaban2019staingan} uses a CycleGAN to map between domains in the MITOS-ATYPIA dataset \citep{roux2014mitos} containing pairs of tissue images where each pair corresponds to images of the same tissue specimen that have been captured using different digital slide scanners. When measuring structural similarity (SSIM) between synthetically transferred images and the corresponding ground truth image, StainGAN was found to be more effective than many traditional methods such as those presented in \citep{reinhard2001color, Macenko2009, khan2014nonlinear, vahadane2016structure}. This demonstrates that CycleGANs are effective at transferring the colour appearance of images from different scanners. Another CycleGAN-based model called Transitive Adversarial Networks (TAN) \citep{Cai2019} proposed an extension of StainGAN that used a more efficient generator and was able to map between domains in the MITOS-ATYPIA dataset more accurately whilst using much less time to process each sample. CycleGANs were augmented with self-attention in \citep{zhang2019self} and a structural consistency loss in the Self-Attentive Adversarial Stain Normalisation (SAASN)\citep{shrivastava2019self} model was found to outperform StainGAN on a large custom dataset of duodenal biopsy slides which originated from one of three sites with significant variation in stain colour distribution across sites.

The DCAE \citep{moyes2018novel} was shown to be effective at learning a normalised representation of digital images of tissue slides that had been captured using two different scanners. However, this learning process required the same tissue specimen be captured using both scanners and then spatially aligned. Such a dataset is expensive and time consuming to produce. In this work, the DCAE model is adapted to model three domains, thereby becoming the MCAE model and its efficacy is evaluated on a novel synthetic dataset. This dataset is generated using a CycleGAN with self-attention; however, unlike many existing methods that use data with relatively low colour variance between domains, the source domains for this dataset are selected to have very high variance between domains. This is beneficial because it enables the evaluation of the MCAE model across a wider variety of tissue image appearances and thus provides insight into the performance of the model on novel data.

\section{CycleGAN Generated Synthetic Dataset}
\label{sec:cyclegan_generated_synthetic_dataset}
Training and evaluating models that are invariant to the effects of staining and scanning becomes much easier when the problem of image alignment can be mitigated. By using cycle-consistent generative adversarial networks (CycleGANs), tissue images from one domain can be mapped to another whilst the underlying tissue structure remains approximately the same.

This section discusses a method for generating a dataset that is aligned by default using self-attentive cycle-consistent adversarial networks. A brief introduction of CycleGANs will be presented in Section \ref{sec:Related_Work_CycleGANs}, followed by a description of the different data domains used to train the models in Section \ref{sec:data_domains}. The architecture and training procedure of the models will be discussed in Sections \ref{sec:cyclegan_architecture} and \ref{sec:CycleGAN_training} respectively. Finally, the approach used to generate the synthetic data points is discussed in Section \ref{sec:data_generation}.

\subsection{Cycle-consistent Adversarial Networks}
\label{sec:Related_Work_CycleGANs}
The goal of a CycleGAN is to map an image from one domain to another, for example: horses to zebras, summer to winter or art to photos \citep{zhu2017unpaired}. This is particularly advantageous in the context of producing a dataset which contains images of the tissue from multiple staining conditions. 

CycleGANs typically contain four primary components: a discriminator for domain A, a discriminator for domain B, a generator for domain A and a generator for domain B. Much like the traditional GAN setup described in \citep{goodfellow2014generative}, the discriminator for domain A learns to distinguish real images from domain A from false/imposter images sampled from the corresponding generator function. The generator for domain A learns to generate realistic false images that appear to belong to domain A, however unlike traditional GAN generators that condition samples on latent variables, the CycleGAN generator conditions the new samples on real observed images from domain B.

In the context of this work, a domain will refer to images that have similar appearance characteristics, such as those created using the same staining and scanning protocols and methods. Therefore, mapping from domain A to domain B refers to the process of adjusting the appearance of images in domain A to match the appearance of those in domain B without affecting the underlying tissue structure.

\subsection{Data Domains}
\label{sec:data_domains}
The proposed generated dataset aims to depict the same tissue from a variety of staining and scanning conditions. Therefore the input data used to train the CycleGAN models should be from a variety of sources. In this section, a brief description of these data sources is presented.

\begin{figure}[h]
    \centering
    \subcaptionbox{}{
        \includegraphics[width=0.2\linewidth]{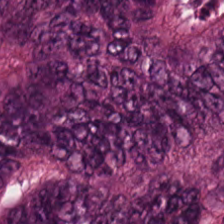}{}%
    }
    \hfill
    \subcaptionbox{}{
        \includegraphics[width=0.2\linewidth]{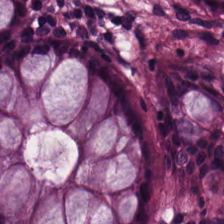}{}%
    }
    \hfill
    \subcaptionbox{}{
        \includegraphics[width=0.2\linewidth]{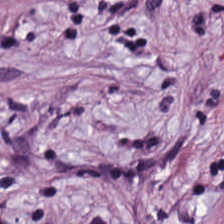}{}%
    }
    \caption{Example 128$\times$128 images from the NCT-CRC dataset}
    \label{fig:Kather_examples}
\end{figure}

\paragraph{NCT-CRC}
This dataset \citep{kather_jakob_nikolas_2018_1214456} contains 100,000 non-overlapping tissue image patches stained with hematoxylin and eosin. The dataset is derived from 86 colorectal tissue slides captured at 20$\times$ resolution that have been split into 224$\times$224 pixel patches and contains a variety of healthy and cancerous tissue. Patches have been divided into a number of anatomical classes, for example, adipose, background, mucosa, smooth muscle and tumour. There are two versions of this dataset, one in which the slides have been normalised using Macenko's slide colour normalisation method \citep{Macenko2009} and the other which has not. In this study the dataset with normalisation is omitted in order to preserve as much of the natural variance as possible. Figure \ref{fig:Kather_examples} shows example patches from this dataset. The desired output size of the generated dataset is 128$\times$128 pixels, thus to transform this dataset to the correct size without affecting the magnification; a 128$\times$128 pixel patch is extracted from the centre of each image.

\begin{figure}[h]
    \centering
    \subcaptionbox{}{
        \includegraphics[width=0.2\linewidth]{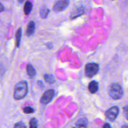}{}%
    }
    \hfill
    \subcaptionbox{}{
        \includegraphics[width=0.2\linewidth]{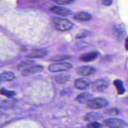}{}%
    }
    \hfill
    \subcaptionbox{}{
        \includegraphics[width=0.2\linewidth]{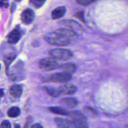}{}%
    }
    \caption{Example 128$\times$128 images from the TCGA-COAD dataset}
    \label{fig:TCGA_COAD_examples}
\end{figure}

\paragraph{TCGA-COAD}
The Cancer-Genome Atlas (TCGA) is a repository of cancer-related data \citep{TCGA_Website}. This includes thousands of digital tissue slides from a variety of anatomical locations and of varying diseases. In order to match the approximate anatomical location of the NCT-CRC dataset, the TCGA-COAD project is selected for these purposes as it contains colorectal tissue images under a variety of staining and scanning conditions as well as varying diagnostic status. A single slide was selected that contained a large proportion of clean, undamaged tissue. This slide was downloaded and divided into approximately 20,000 non-overlapping patches, each 128$\times$128 pixels in size. The slide was captured at 20$\times$ resolution and was selected by manually comparing the colour appearances with the NCT-CRC dataset images and attempting to maximise the differences. Figure \ref{fig:TCGA_COAD_examples} shows examples of patches derived from the selected TCGA-COAD slides.

\begin{figure}[h]
    \centering
    \subcaptionbox{}{
        \includegraphics[width=0.2\linewidth]{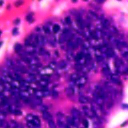}{}%
    }
    \hfill
    \subcaptionbox{}{
        \includegraphics[width=0.2\linewidth]{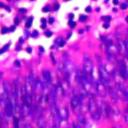}{}%
    }
    \hfill
    \subcaptionbox{}{
        \includegraphics[width=0.2\linewidth]{sections/images/generated_dataset/Nantong/b.png}{}%
    }
    \caption{Example 128$\times$128 images from the NTSPH dataset}
    \label{fig:Nantong_examples}
\end{figure}

\paragraph{NTSPH Dataset}
The third dataset used in this study is sourced collection from the Second People's Hospital of Nantong (NTSPH), China. It contains 18 hematoxylin and eosin stained tissue images of colorectal tissue captured at 20$\times$ resolution. The tissue in this dataset demonstrates very strong staining characteristics as can be seen in Figure \ref{fig:Nantong_examples} alongside high average brightness. Although the tissue type is the same as the NCT-CRC and TCGA-COAD datasets, the variance in staining and capture appearances is large, making this dataset very useful for the purposes of capturing and normalising tissue-image colour variation.

Figures \ref{fig:Kather_examples} - \ref{fig:Nantong_examples} highlight the diverse colour spectra found across the hematoxylin and eosin stained images from different centres. The NCT-CRC dataset heavily features dark purples and reds/purple whilst in contrast the NTSPH dataset is primarily comprised of bright pinks and blue/purple pixels. The colour spectra of the TCGA-COAD data are in stark contrast again featuring more light blues and pinks. Despite all three datasets consisting of colorectal tissue, stained with hematoxylin and eosin, and extracted at the same magnification resolution, the resulting images are distinct to each dataset.  

\subsection{CycleGAN Architecture}
\label{sec:cyclegan_architecture}
\begin{figure}[h]
    \centering
    \includegraphics[width=1.0\linewidth]{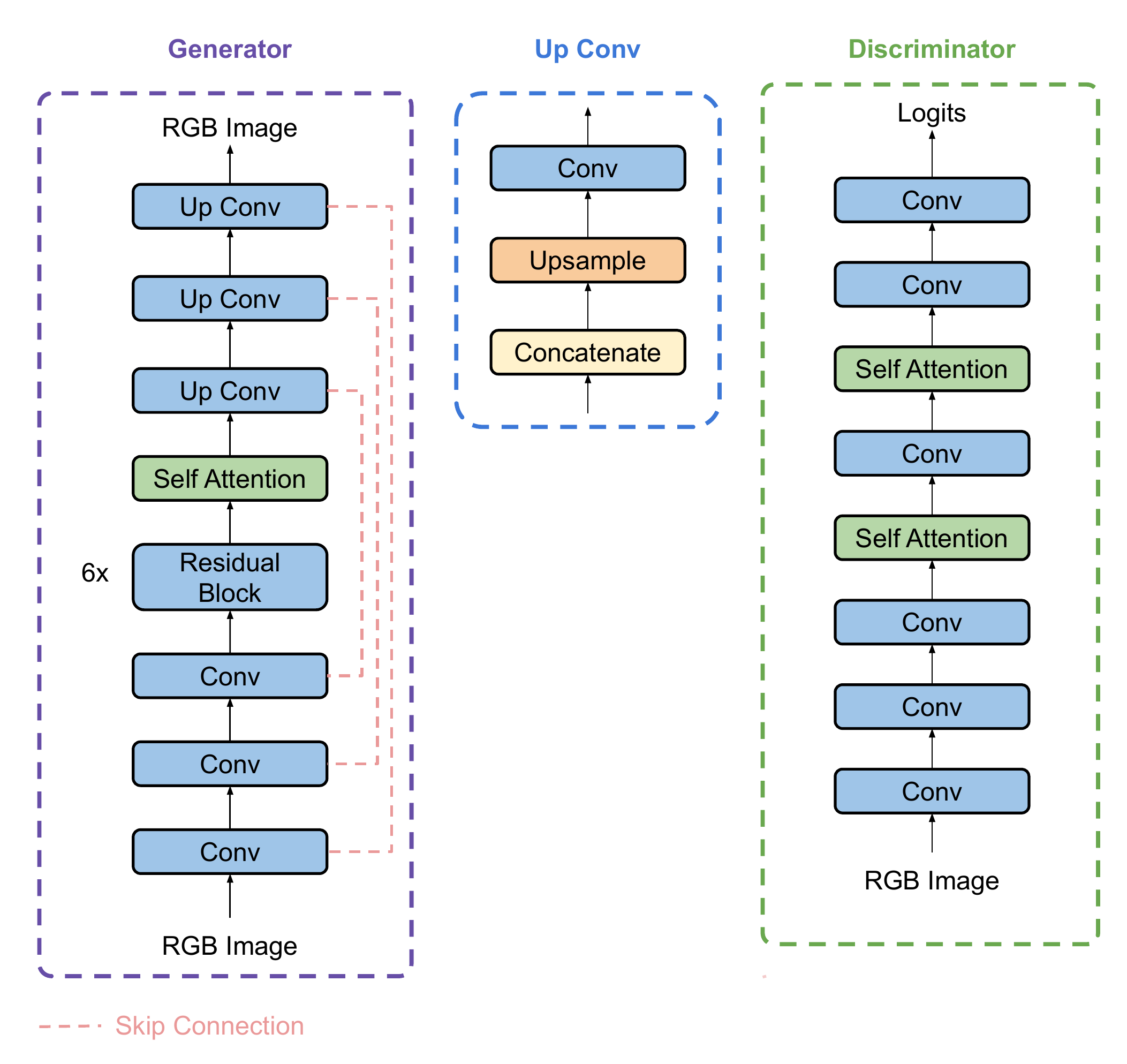}
    \caption{CycleGAN model architectures.}
    \label{fig:CycleGAN_Architectures}
\end{figure}

The architecture used to generate the synthetic dataset follows the structure defined in Self-Attentive Adversarial Stain Transfer model \citep{shrivastava2019self}. This model is a CycleGAN structure similar to that of StainGAN \citep{shaban2019staingan}; however each convolutional layer in the generator and discriminator networks are followed by a self-attention layer \citep{zhang2019self}. The self-attention element provides less reliance on local receptive field of neurons than traditional GANs and enables long-range dependencies to be learnt. The self-attention creates a balance between the model's ability to capture long-range dependencies whilst maintaining computational and statistical efficiency. For this work, the generator and discriminator architectures are adjusted to improve performance but in general resemble those defined in \citep{shrivastava2019self}. Figure \ref{fig:CycleGAN_Architectures} depicts the architectures used for the generator and discriminator portions of the CycleGAN.

The generator architecture follows the U-Net encoder-decoder pattern with skip connections between corresponding layers. Input images of size 128$\times$128 pixels are downsampled by three convolutional layers, each of which are followed by spectral normalisation and the ReLU activation function. This results in a 16$\times$16$\times$256 feature map that is subsequently processed by 6 residual layers \citep{he2016deep}, leaving the spatial and channel dimensionality of the feature maps the same. Finally, three transpose convolutional layers upsample the feature maps, doubling the spatial dimensionality at each layer. The final output of the generator is an RGB image of 128$\times$128 pixels in size.

The discriminator model receives 128$\times$128 pixel RGB images as input. The discriminator is comprised of 6 convolutional layers each with the ReLU activation function and spectral normalisation. Self-attention layers are added after the third and fourth convolutional layers as these have relatively low spatial dimensional (16$\times$16 pixels).

For the work in this paper, these models are slightly adjusted to improve memory efficiency by only adding self-attention layers after convolutional layers with relatively small spatial dimensions. This still allows for the benefits of capturing the long-range spatial dependencies via the self-attention layers but avoids the large memory costs incurred when applying self-attention to feature maps with large spatial dimensionality.

\subsection{CycleGAN training}
\label{sec:CycleGAN_training}
The goal of the model training is to learn two mapping functions, $F$ and $G$ to associate two domains $A$ and $B$ such that $F : A \mapsto B$ and $G : B \mapsto A$. Domains $A$ and $B$ may refer to images from the TCGA-COAD and NCT-CRC datasets discussed in Section \ref{sec:data_domains} respectively. To learn these functions, a cyclical procedure is defined that combines three loss functions: the GAN loss, the identity loss and the cycle consistency loss.

\paragraph{GAN Loss} A real image from domain $A$ is mapped to domain $B$ using the function $F$ to produce a false image $\hat{B} = F(A)$, likewise a real image from domain $B$ is mapped to domain $A$ using the function $G$ to produce a counterfeit (or fake) image $\hat{A} = G(B)$. The GAN loss uses each domain's discriminator to determine how closely the fake images match their new domain. The GAN loss for the mapping $A \mapsto B$ is defined in Equation \ref{eq:generated_dataset:GAN_loss_F}.

\begin{equation}
    \begin{split}
    \ell_{GAN}(F, D_B, A, B) & = \mathbb{E}_{b \backsim p_{data}(B)} [log D_B(b)] \\
    & + \mathbb{E}_{a \backsim p_{data}(A)}[log(1 - D_B(F(a)))]
    \end{split}
    \label{eq:generated_dataset:GAN_loss_F}
\end{equation}

where $F$ is the generative function mapping images from domain A to domain B, $D_B$ is the discriminative function which predicts whether a sample image belongs to domain B or is an imposter, $p_{data}(A)$ represents the data-generating distribution of domain $A$ and $a \backsim p_{data}(A)$ represents a real image sampled from this data-generating distribution. Therefore, $\mathbb{E}_{a \backsim p_{data}(A)}$ represents the expected value over the distribution of all real images $(a)$ from the data-generating distribution of domain $A$. Similarly, $p_{data}(B)$ represents the data-generating distribution of domain $B$, $b \backsim p_{data}(B)$ represents a real image sampled from this other data-generating distribution and $\mathbb{E}_{b \backsim p_{data}(B)}$ represents the expected value over the distribution of all real images $(b)$ from the data-generating distribution of domain $B$. $F(a)$ represents a real image from domain A that has been mapped to domain B by the generative function $F$. Likewise, the GAN loss for the mapping $B \mapsto A$ is defined in Equation \ref{eq:generated_dataset:GAN_loss_G}.

\begin{equation}
    \label{eq:generated_dataset:GAN_loss_G}
    \begin{split}
    \ell_{GAN}(G, D_A, A, B) & = \mathbb{E}_{a \backsim p_{data}(A)}[log D_A(a)] \\
    & + \mathbb{E}_{b \backsim p_{data}(B)}[log(1 - D_A(G(b)))]
    \end{split}
\end{equation}

where $G$ is the generative function mapping images from domain B to domain A and $D_A$ is the discriminative function that predicts where a sample image belongs to domain A or is an imposter.

These equations define a min-max objective function where the generator ($G$ or $F$) is tasked with minimising the function whilst its discriminative counterpart ($D_A$ or $D_B$) aims to maximise it. Taking the mapping $A \mapsto B$ as an example, to maximise this function the discriminator must learn a decision boundary between real and fake images by assigning high scores to real images (thus $logD_A(a) \to 0$) and assigning low scores to false images (thus $log(1 - D_A(G(b))) \to 0$). The only way the generator is able to minimise the objective function is by ensuring the discriminator assigns high scores to the fake images it produces (thus $log(1 - D_A(G(b))) \to -\infty$). The equivalent principle holds for the mapping $B \mapsto A$.

\paragraph{Identity loss} It has been shown in \citep{taigman2016unsupervised} that adding the identity loss improves the colour and tint mapping of translated images. This loss function for both domains is defined as below:

\begin{equation}
\label{eq:generated_dataset:identity_loss}
\begin{split}
    \ell_{identity}(F, G, A, B) & = \mathbb{E}_{a \backsim p_{data}(A)}[\|G(a) - a\|_1] \\ 
    & + \mathbb{E}_{b \backsim p_{data}(B)}[\|F(b) - b\|_1]
\end{split}
\end{equation}

where $a \backsim p_{data}(A)$ is a real image sampled from the data distributution of domain $A$ and similarly $b \backsim p_{data}(B)$ is a real image sampled from the data distribution of domain $B$. $\mathbb{E}_{a \backsim p_{data}(A)}[\|G(a) - a\|_1]$  represents the reconstruction loss averaged across all real samples.

\paragraph{Cycle consistency loss} A generator with large enough capacity may be able to map an image from the source domain to that of a target domain, such that the output distribution of the generator matches the data-generating distribution of the target domain. However because CycleGANs are trained with unaligned datasets, a GAN loss alone is not enough to ensure that the mapped images maintain resemblance across domains. For example, if a CycleGAN has learned a mapping between photographs of people and paintings, it would be expected that the mapping from a photograph of person A would resemble a painting of person A rather than a painting of person B or person C. Using only a GAN loss, a mapping from a photograph of person A to a painting of person B would be valid, despite being undesirable. The introduction of the cycle-consistency loss has been shown to enhance resemblance across domains. To calculate this loss, a real image from a source domain $a \backsim p_{data}(A)$ is mapped to to a target domain, $\hat{b} = F(a)$. This generated image in the target domain is then mapped back to the source domain using the appropriate generating function $\hat{a} = G(\hat{b})$. An $\ell_1$ loss is then calculated between the twice-mapped image and the original input as described in Equation \ref{eq:generated_dataset:cycle_consistency}.

\begin{equation}
\label{eq:generated_dataset:cycle_consistency}
\begin{split}
    \ell_{cyc}(F, G, A, B) & = \mathbb{E}_{a \backsim p_{data}(A)}[\| G(F(a)) - a\|_1] \\
    & + \mathbb{E}_{b \backsim p_{data}(B)}[\| F(G(b)) - b \|_1]
\end{split}
\end{equation}

\paragraph{Full objective} By combining the two GAN losses (for $F: A \mapsto B$ and $G: B \mapsto A$), the identity loss and the cycle-consistency loss, we arrive at the final objective function defined in Equation \ref{eq:generated_dataset:full_objective}.

\begin{equation}
\label{eq:generated_dataset:full_objective}
\begin{split}
    \ell(G, F, D_A, D_B) & = \ell_{GAN}(F, D_B, A, B) \\
    & + \ell_{GAN}(G, D_A, A, B) \\
    & + \lambda_1 \ell_{identity}(F, G, A, B) \\
    & + \lambda_2 \ell_{cyc}(F, G, A, B)
\end{split}
\end{equation}

where $\lambda_1$ and $\lambda_2$ control the relative contribution of the identity and cycle-consistency losses.

\subsubsection{CycleGAN optimisation procedure}
Given the full objective function defined in Equation \ref{eq:generated_dataset:full_objective}, the optimisation problem can be defined as:

\begin{equation}
\label{eq:generated_dataset:problem}
G^*, F^* = arg \min_{G,F} \max_{D_A, D_B} \ell(G, F, D_A, D_B)
\end{equation}

The optimal generators and discriminators are learned by alternating between updating the parameters of the generators, followed by updating the parameters of the discriminators. For the generators, a full cycle is performed as described in the following 4 steps: 

\begin{enumerate}
    \item The source domain is mapped back to itself using the target generator (e.g. $G: B \mapsto A, \hat{a} = G(a)$) and the identity loss is calculated (Equation \ref{eq:generated_dataset:identity_loss})
    \item A minibatch of real source images are mapped to the target dataset using the appropriate generator. The GAN loss is calculated at this stage (Equations \ref{eq:generated_dataset:GAN_loss_F} \& \ref{eq:generated_dataset:GAN_loss_G})
    \item The generated images are then mapped back to the source domain using the complementary generator and the cycle-consistency loss is calculated (Equation \ref{eq:generated_dataset:cycle_consistency})
    \item The losses are weighted and summed. Gradients are calculated using backpropagation and the parameters are updated via the Adam optimisation algorithm.
\end{enumerate}

\subsection{Data Generation}
\label{sec:data_generation}
\begin{figure}
    \centering
    \includegraphics[width=0.9\linewidth]{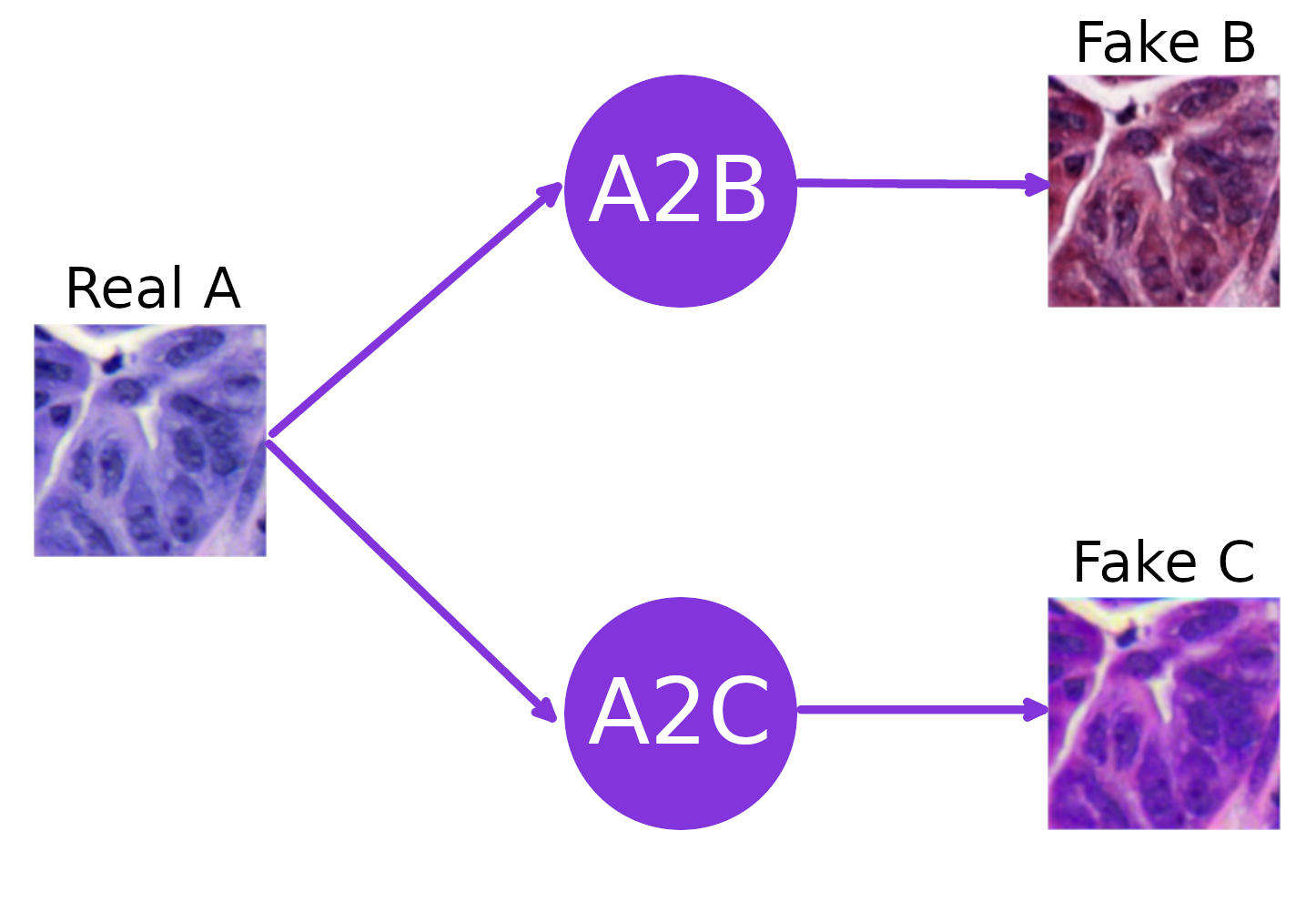}
    \caption{Triplet generation: Appropriate generators from each CycleGAN model are selected (depicted as purple circles) and used to map real patches from TCGA-COAD domain (A) to domains for the NCT-CRC dataset (B) and NTSPH dataset (C).}\label{fig:triplet_generation}
\end{figure}

\begin{figure}   
    \centering
    \subcaptionbox{}{
        \includegraphics[width=0.2\linewidth]{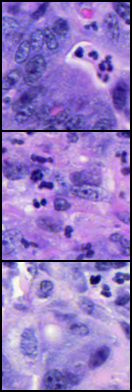}{}%
    }
    \hfill
    \subcaptionbox{}{
        \includegraphics[width=0.2\linewidth]{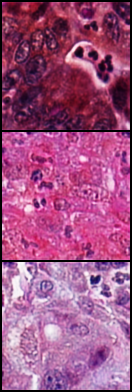}{}%
    }
    \hfill
    \subcaptionbox{}{
        \includegraphics[width=0.2\linewidth]{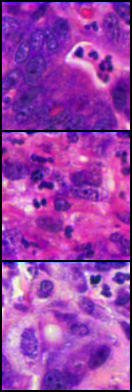}{}%
    }
    \caption{Example triplets from the synthetic dataset. (a) real images from TCGA-COAD, (b) synthetic images (TCGA-COAD $\mapsto$ NCT-CRC), (c) synthetic images (TCGA $\mapsto$ NTSPH)}
    \label{fig:triplet_examples}
\end{figure}

The final dataset $X$ produced by the CycleGAN is comprised of 20,000 triplets of tissue-image patches. Each triplet represents the same tissue structure but with the colour appearance of the domains discussed in Section \ref{sec:data_domains}. To achieve this, two CycleGAN models are trained independently; the first mapping between the domains of the TCGA-COAD and NCT-CRC datasets and the second mapping between the domains of the TCGA-COAD and NTSPH datasets. Once both models are trained following the processes described in Section \ref{sec:CycleGAN_training}, we have a mechanism to map between data from the domain of the TCGA-COAD dataset and both of the domains of the other datasets. Given the trained models, the dataset is generated by mapping every real image patch in the TCGA-COAD dataset onto the domains of the NCT-CRC and NTSPH datasets using the appropriate generator from each CycleGAN model. Figure \ref{fig:triplet_generation} illustrates the process of generating the dataset. Figure \ref{fig:triplet_examples} shows example triplets in the dataset generated through the process illustrated in Figure \ref{fig:triplet_generation}. Any image patch in $X$ can be referenced using a triplet index $i$ and a domain index $d \in \{A, B, C\}$. For example, $X_{i,A}$ refers to the image patch for domain A in the $i_{th}$ triplet.

\section{Multi-Channel Auto-Encoder}
This section will outlined the methodology behind the proposed Multi-Channel Auto-Encoder model and how it relates to the synthetic dataset established in Section.
\label{sec:multi_channel_auto_encoder}
The core idea of the Multi-Channel Auto-Encoder model and Dual-Channel Auto-Encoder model \citep{moyes2018novel} is that there exists some representation of histopathology images where some change in the staining or scanning conditions of an image does not correspond to a change in the representation of the images, i.e. the representation is invariant to scanning and staining parameters. It was shown in \citep{moyes2018novel} that the DCAE model is able to learn a representation of histopathology images that is more invariant to scanner-induced colour variance compared to a similar method called StaNoSA \citep{janowczyk2017stain}. This was achieved by training a series of auto-encoders, one for each domain and using a combination of a feature similarity loss and a cluster-based loss to maximise similarity between the features of image-patches representing the same tissue from each of the encoders. In this section, a brief outline of the Multi-Channel Auto-Encoder model is given but more details are available in \citep{moyes2018novel}. The MCAE model can be decomposed into a set of encoders $F = \{f_A, f_B, f_C\}$ and a set of decoders $G = \{g_A, g_B, g_C\}$.

Following the same process as the DCAE model, the encoding process for the MCAE model involves dividing the $i^{th}$ image patch for domain $d$, $X_{i,d}$ into a set of overlapping 8x8 sub-patches and applying the encoding function $f_d$ to each sub-patch. The encoding function maps each 8x8 sub-patch into a 10-dimensional feature vector. For the $i^{th}$ triplet $X_i$, the tensor $\mathbf{Z} \in \mathbb{R}^{3 \times J \times 10}$ represents the encoded feature representation of each image the triplet, where 3 corresponds to domain A, B or C, $J$ is the total number of 8x8 patches that can be extracted from each image and 10 is the size of the feature vectors. A reconstruction of $X_{i,d}$ is formed by applying the decoder function $g_d$ to every feature vector in $\mathbf{Z}_{d,:,:}$.

\subsection{Model Architecture}
Given $N$ disjoint data domains, the MCAE model requires $N$ auto-encoders. These auto-encoders process an image in a patch-by-patch fashion and each of these patches are typically 8x8 pixels. Each 8x8 pixel patch is mapped onto a 10-dimensional feature vector by the encoder and this feature vector is mapped to a reconstruction of the 8x8 pixel patch by the decoder. Each encoder is a multi-layer perceptron with 192 nodes in the input layer (8x8 pixels and 3 colour channels), 100 neurons in the hidden layer and 10 neurons in the output layer. Each neuron in the hidden and output layers use the hyperbolic tangent activation function. The decoders have a symmetrical architecture, with 10 nodes in the input layers, 100 neurons in the hidden layer and 192 neurons in the output layer; however the sigmoid activation function is used on the output of the decoder in order to map back to the range of the RGB space. The specific architecture used for the MCAE model is chosen so as to remain as close as possible to previous work in \citep{moyes2018novel} and \citep{janowczyk2017stain}.

\subsection{Reconstruction Loss}
\label{sec:reconstruction_loss}
The primary training signal of the MCAE model is the reconstruction loss. For any 8x8 pixel sub-patch, the reconstruction loss quantifies the error between the original sub-patch and the reconstruction of it formed by the auto-encoder. Minimising this reconstruction loss ensures that the features produced by each of the encoding functions in $F$ are information-rich. This is because the encoder functions act as a form of compression while the decoders act as decompression. The only way for the decoders to accurately reconstruct the original sub-patch is if the feature vector has captured key information about the sub-patch.

\subsection{Feature Loss}
\label{sec:feature_loss}
Given that each image patch in the $i^{th}$ triplet represents the same tissue with different colour characteristics, it follows that this is also true for each of the 8x8 sub-patches. In order to learn a representation of images that is invariant to changes in colour characteristics, a feature loss is used. This feature loss quantifies the differences between the encoded feature representations of sub-patches using the mean-squared error. For the $i^{th}$ triplet, this is calculated as per Equation \ref{eq:feature_loss}.

\begin{equation}
    L_{feature} = \frac{1}{2 \times 10 \times J}  \sum_{j}^{J} \sum_{z}^{10} (Z_{A,j,z} - Z_{B,j,z})^2 + (Z_{A,j,z} - Z_{C,j,z})^2
    \label{eq:feature_loss}
\end{equation}

Minimising this loss function alongside the reconstruction loss ensures that the encoded representations in $\mathbf{Z}$ are feature rich (due to minimisation of the reconstruction loss) and colour appearance invariant (due to the feature loss). 

\subsection{Cluster Loss}
\label{sec:cluster_loss}
The center loss \citep{wen2016discriminative} has been shown to improve the performance of classification algorithms by bringing the feature representations of separate instances of the same class closer to each other than to the feature representations of instances of other classes. This is problematic in this work as there are no class labels to utilise; therefore these labels are approximated through pseudo-labels that are derived from a K-Means clustering algorithm that has been fitted on randomly sampled feature vectors from patches of images in domain A (TCGA-COAD) and is re-fitted at the end of each training epoch. The cluster loss can be calculated as per Equation \ref{eq:cluster_loss}
\begin{equation}
    \begin{split}
        L_{cluster} & = \| Z_{A, i, j} - \mu_{i,j} \|_2^2 \\
        & + \| Z_{B, i, j} - \mu_{i,j} \|_2^2 \\
        & + \| Z_{C, i, j} - \mu_{i,j} \|_2^2
    \end{split}
    \label{eq:cluster_loss}
\end{equation}
where $\mu_{i,j}$ is the centroid corresponding to the K-Means determined pseudo-label for $Z_{A,i,j}$ that represents the encoded feature representation of the $j^{th}$ patch of the image for domain A in the $i^{th}$ triplet, and so on for $Z_{B,i,j}$ and $Z_{C,i,j}$.

Both the feature loss defined in Section \ref{sec:feature_loss} and the cluster loss defined in this section behave similarly, in that they are both explicitly influencing the feature representations of certain patches. The feature loss brings the feature representations of patches that are spatially equivalent within triplets closer together. The cluster loss brings the feature representations of patches that are categorically equivalent closer together. Due to the cluster loss being based on the average of many similar features vectors of patches belonging to the same class, it presents a slower moving, more robust direction in which to move when compared to the feature loss.

\subsection{Combined Objectives}
The overall objective function of the MCAE model is defined through a summation of the reconstruction loss defined in Section \ref{sec:reconstruction_loss}, the feature loss defined in Section \ref{sec:feature_loss} and the cluster loss defined in Section \ref{sec:cluster_loss}.
\begin{equation}
    \argmin_W L = L_{reconstruction} + L_{feature} + L_{cluster}
\end{equation}
where $W$ is the set of all model parameters for all auto-encoders.

\section{Experimental Evaluation}\label{sec:experimental_eval}
The experimental evaluation in this work is split into three parts. In Section \ref{sec:synthetic_dataset_evaluation}, the quality of the CycleGAN-based colour normalisation approach is evaluated in order to determine its validity for use in subsequent experiments. Next, in Section \ref{sec:domain_invariance}, the similarity between learned feature representations is compared for the MCAE and StaNoSA models. Finally, in Section \ref{sec:tissue_classification}, the quality of these learned feature representations is evaluated on a classification task using the NCT-CRC dataset \cite{kather_jakob_nikolas_2018_1214456}.
\subsection{Synthetic Dataset Evaluation}
\label{sec:synthetic_dataset_evaluation}
In this section, the CycleGAN-generated, synthetic dataset will be evaluated with respect to the quality of colour normalisation between the domains discussed in Section \ref{sec:data_domains}.

Many existing methods that make use of CycleGAN-based models evaluate their approaches on the MITOS-ATYPIA dataset because it contains images of the same tissue specimens captured by different digital slide scanners. However, the colour distributions between domains in the MITOS-ATYPIA dataset are relatively similar. In this work, the effectiveness of CycleGAN-based models is demonstrated on data with much larger variance in colour distributions due to significant variations in the staining and scanning processes. Previous work \citep{bejnordi2015stain} has used the normalised median intensity (NMI) to measure colour similarity before and after colour normalisation, however this approach requires the use of nuclei segmentation masks; which are not available in this work. Instead, similarity measures based on the hue-saturation-density (HSD) space are used. The HSD transform is the hue-saturation-intensity (HSI) colour transformation applied to optical density images rather than RGB images. Much like the HSI colour space, the HSD colour space transforms tissue images into two colour components - hue and saturation, as well as a density component which represents the underlying tissue structure. The chromatic components of the HSD space are defined by the $c_x$ and $c_y$ attributes, from which the hue and saturation values are derived. The $c_x$ component for each pixel is calculated as per Equation \ref{eq:HSD_cx}.
\begin{equation}
    c_x = \frac{I_R}{I} - 1
    \label{eq:HSD_cx}
\end{equation}
where $I_R$ is the red optical density value at the pixel and $I$ is the average optical density value of the pixel across each colour channel (R, G, and B). The $c_y$ component for each pixel can then be calculated as per Equation \ref{eq:HSD_cy}.
\begin{equation}
    c_y = \frac{I_G - I_B}{\sqrt{3} \cdot I}
    \label{eq:HSD_cy}
\end{equation}
where $I_G$ is the pixel-wise green optical density value, $I_B$ is the pixel-wise blue optical density value and $I$ is the pixel-wise average optical density value.

The HSD colour space \citep{van2000hue} has been used in existing colour normalisation algorithms \citep{geijs2018automatic, bejnordi2014quantitative} where the density channel is fixed and the hue and saturation values are transformed across domains. Therefore the HSD space provides the opportunity for evaluating colour normalisation quality. The HSD-based evaluation strategy is based on the following two criteria:

\begin{enumerate}
    \item Colour normalisation: a mapping from domain A to domain B will be deemed as successful if the $c_x$ and $c_y$ components (corresponding to hue and saturation) of the synthetic images of domain B (i.e. images from domain A that have been normalised using the CycleGAN generator) match the components of the real images from domain B.
    \item Density similarity: throughout the mapping process, the underlying tissue structure of images should remain the same. If the density components of synthetic images roughly match those of the real images, then it can be assumed that the tissue structure has been preserved.
\end{enumerate}

\subsubsection{Colour Normalisation}
The quality of colour normalisation achieved by the CycleGAN is determined through analysis of the $c_x$ and $c_y$ components of the HSD space corresponding to hue and saturation respectively. As discussed in Section \ref{sec:synthetic_dataset_evaluation}, the CycleGAN-based colour normalisation from a source domain (e.g. the domain of the TCGA-COAD dataset) to a target domain (e.g. the domain of the NCT-CRC dataset) will be deemed successful if the distribution of hue and saturation components for the synthetic output images (e.g. the domain of the TCGA-COAD dataset $\mapsto$ the domain of the NCT-CRC dataset via CycleGAN generator) matches the real target domain more than the real source domain.

\begin{figure}[h!]
\centering
\begin{subfigure}[b]{0.95\linewidth}
    \centering
    \includegraphics[width=0.95\linewidth]{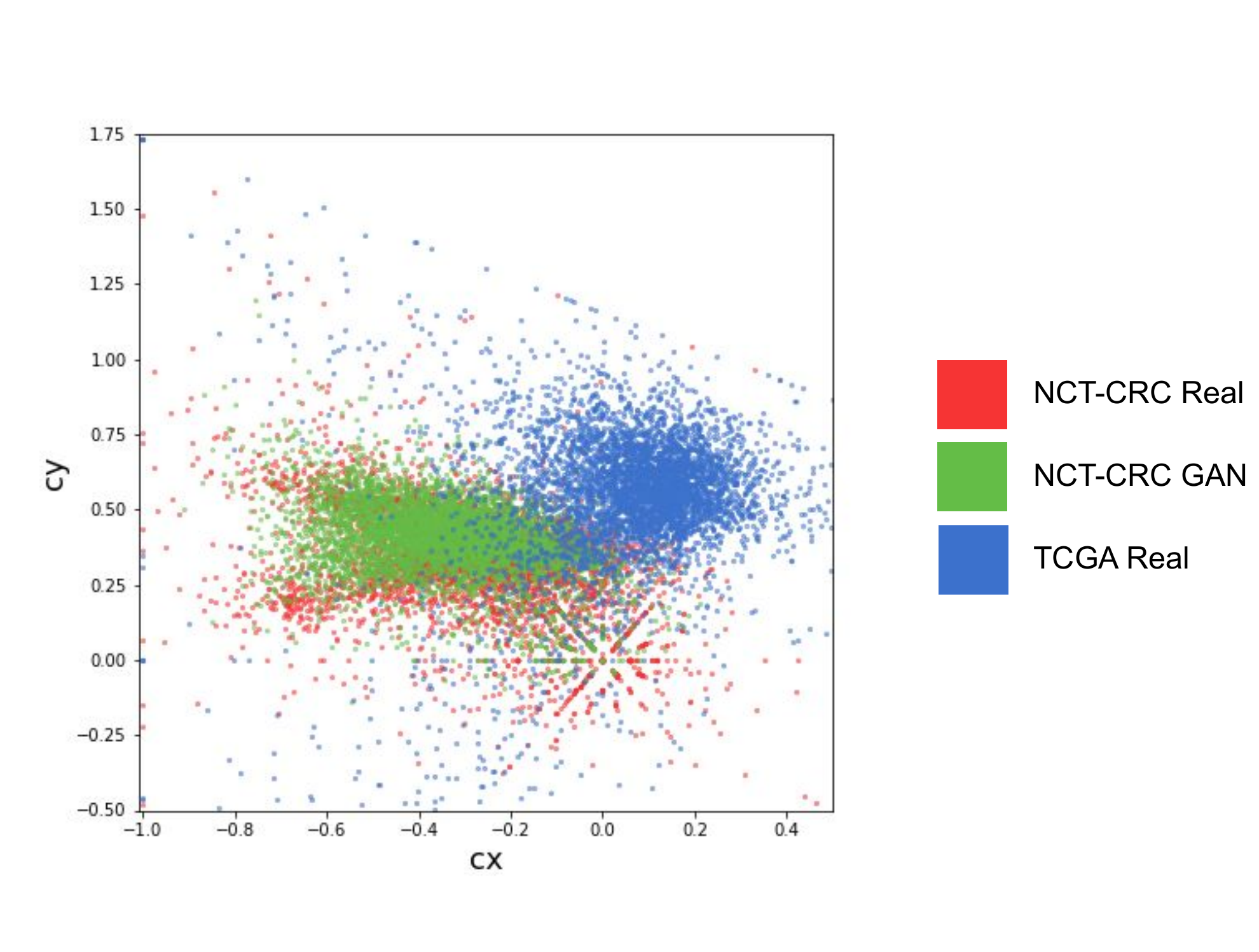}
    \caption{$c_x$ and $c_y$ components of pixels randomly sampled from real images in the from the TCGA-COAD dataset (blue), real images in the NCT-CRC dataset (red) and synthetic images that have been mapped from the domain of the TCGA-COAD dataset to the domain of the NCT-CRC dataset using a CycleGAN (green).}
    \label{fig:cxcy_distributions_kather_tcga}
\end{subfigure}
\begin{subfigure}[b]{0.95\linewidth}
    \centering
    \includegraphics[width=0.95\linewidth]{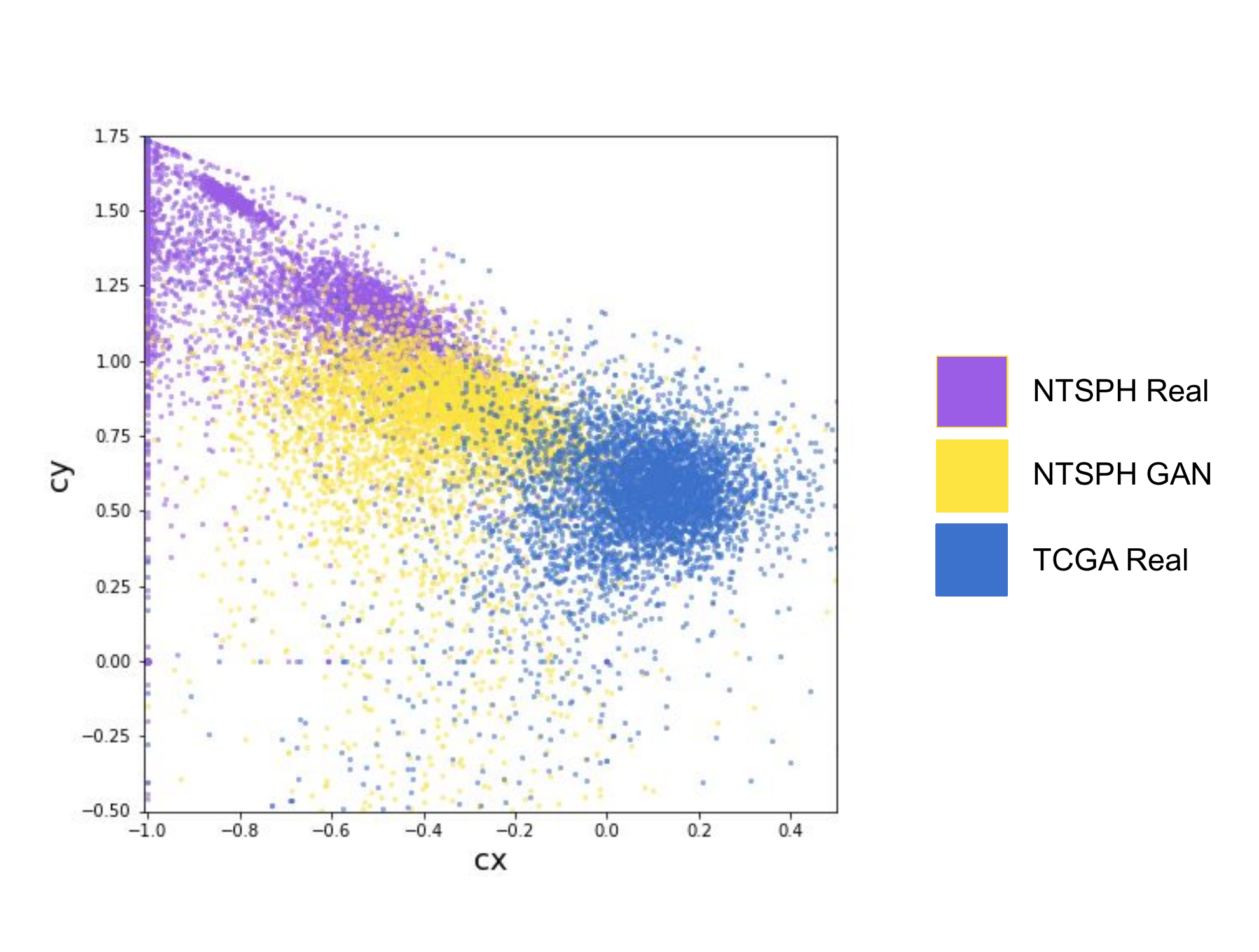}
    \caption{$c_x$ and $c_y$ components of pixels randomly sampled from real images in the TCGA-COAD dataset (blue), real images in the NTSPH dataset (purple) images and synthetic images that been mapped from the domain of the TCGA-COAD dataset to the domain of the NTSPH dataset using a CycleGAN (yellow).}
    \label{fig:cxcy_distributions_nantong_tcga}
\end{subfigure}
\caption{Distributions of $c_x$ and $c_y$ components of the real and synthetic data distributions.}
\label{fig:cxcy_distributions}
\end{figure}

In Figure \ref{fig:cxcy_distributions_kather_tcga}, pixels are randomly sampled from images in the real domains of the NCT-CRC and TCGA datasets and finally images from the synthetic `NCT-CRC GAN' domain. The `NCT-CRC GAN' domain refers to real images from the TCGA domain that have had their appearance normalised to match the domain of the NCT-CRC dataset using the CycleGAN generator. These randomly sampled pixels are plotted in Figure \ref{fig:cxcy_distributions_kather_tcga} according to their corresponding $c_x$ and $c_y$ components. The significant overlap of the pink and purple data-points in Figure \ref{fig:cxcy_distributions_kather_tcga} demonstrate that the distribution of $c_x$ and $c_y$ components in the `NCT-CRC GAN' domain match the `NCT-CRC Real' domain much more closely than the original TCGA-COAD domain. This suggests that the CycleGAN architecture described in Section \ref{sec:cyclegan_architecture}, is able to successfully map the colour appearance of images between domains with very different staining conditions.

A similar approach has been taken to produce Figure \ref{fig:cxcy_distributions_nantong_tcga}. However, in this figure, the pixels are randomly sampled from the real NTSPH domain, the real TCGA domain and finally the synthetic `NTSPH GAN' domain, which similarly refers to images from the real TCGA-COAD dataset that have had their colour appearance mapped using a CycleGAN generator to match the colour appearance of the real NTSPH domain. It can be seen in Figure \ref{fig:cxcy_distributions_nantong_tcga} that the transformation of component distributions from the TCGA to NTSPH domains has not been as successful as with the NCT-CRC domain in Figure \ref{fig:cxcy_distributions_kather_tcga}. Examining Figure \ref{fig:cxcy_distributions_nantong_tcga} shows that the real NTSPH domain is extreme in both $c_x$ and $c_y$ values. This observation concurs with the visual properties of the NTSPH dataset seen in Figure \ref{fig:triplet_examples}. For example, the images from the NTSPH domain are largely pink and purple which reside near an extreme of the unravelled hue line in the HSD space, and therefore have extreme values in the $c_x$ axis. Additionally, the colours observed are `strong` and so have high saturation values seen on the $c_y$ axis. It appears the CycleGAN has struggled to fully translate these values which suggests the model may not be totally sufficient when it comes to the task of normalising tissue image appearance between domains with very large discrepancies in colour distribution. Despite the imperfect mapping of appearance from the TCGA to NTSPH domains, the purpose of this dataset is to present images of the same underlying tissue but with significant variation in their colour appearance. With this in mind, the colour appearance normalisation results appear to be adequate with respect to the hue and saturation components of the HSD space. In Section \ref{sec:experimental_results_density_similarity}, the quality of the normalisation process will be evaluated with respect to the similarity of the density component.

\subsubsection{Density Similarity}
\label{sec:experimental_results_density_similarity}
Within the synthetic dataset, images of the same tissue should differ only by their colour appearance. The degree to which the CycleGAN-based normalisation procedure has altered the underlying tissue structure can be measured by evaluating the similarity of the density components across domains for images of the same tissue. Ideally, the density component would be the same across domains.

Table \ref{tab:density_similarity} shows the average similarity scores of the density channels between domains for structural similarity index (SSIM). In general, the differences in the density channels are relatively low. For example, the SSIM score of 0.852628 between domains A and C suggests a high degree of similarity between the density channels. The SSIM score of 0.818682 between domain A and B suggests there is less similarity between the two domains. These observations are also reflected in the images in Figure 8 where it can be seen the density images are more similar between domains B (NCT-CRC) and C (NTSPH) than between domains A (TCGA-COAD) and B. 
%
%
\begin{table}[h]
\centering
\caption{Mean and standard deviation of SSIM values calculated between the density components of source and target domains within triplets.}
\vspace{1em}
\begin{tabular}{llrr}
\toprule
\hline
source & target & mean & std \\
\midrule
\hline
A & B &  0.818682 &  0.119560 \\
A & C &  0.852628 &  0.047245 \\
B & C &  0.865696 &  0.055421 \\
\bottomrule
\hline
\end{tabular}
\label{tab:density_similarity}
\end{table}

\begin{figure}[h]
    \centering
    \includegraphics[width=\linewidth]{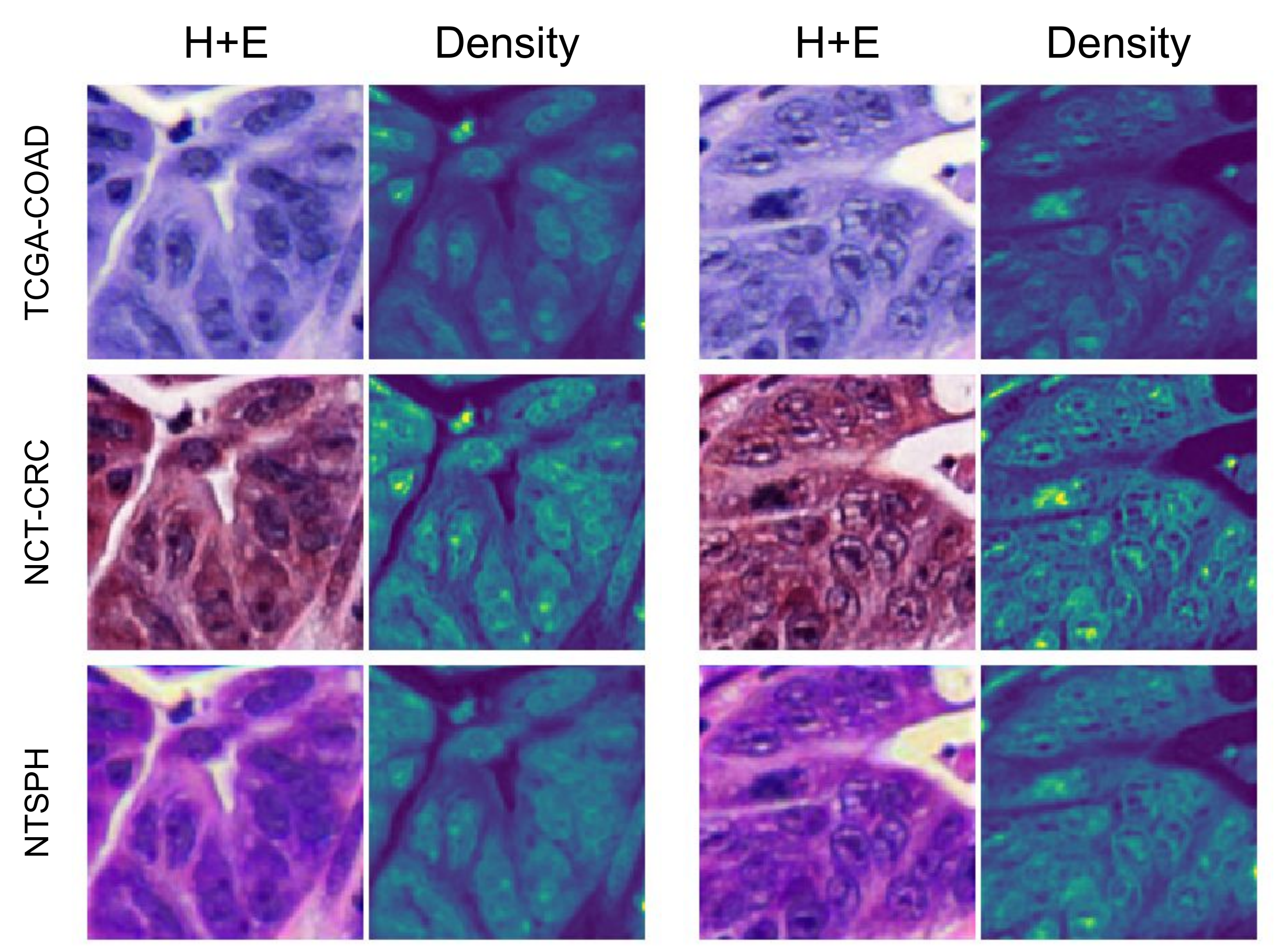}
    \caption{Example triplets from the synthetic dataset (`H+E` columns) and their corresponding density components (`Density` columns). The `TCGA-COAD` images are real, the `NCT-CRC` and `NTSPH` images are synthetically generated using the CycleGAN and the real `TCGA-COAD` image. Visually, the tissue structure appears to have been mostly preserved in the `H+E` images, but the `Density` images suggest that some degradation of tissue structure has occurred during the normalisation process.}
    \label{fig:density_comparison}
\end{figure}

Figure \ref{fig:density_comparison} depicts two real images from the TCGA domain followed by corresponding synthetic images. Each image is accompanied by a visualisation of its density channel. Visually, it can be seen that the density channels have a very similar structure, implying the overall structure is relatively consistent. However, there are some obvious differences in the overall intensity, for example, the synthetic images from the NCT-CRC domain appear brighter than the TCGA and NTSPH domains; especially in areas containing nuclei. This suggests the overall generative process has not entirely preserved the tissue structure. Despite these differences, the overall structure of the density channels remain similar. Combining this with the good colour normalisation results shown in Figure \ref{fig:cxcy_distributions} and the example triplets in Figure \ref{fig:triplet_examples} leads to the conclusion that the CycleGAN-generated synthetic dataset has achieved its primary purpose of depicting the same tissue specimens from a variety of stain colour appearances.
%
%
%
\subsection{Domain Invariance}
\label{sec:domain_invariance}
This section explores the efficacy of the MCAE and StaNoSA models on the CycleGAN-generated synthetic dataset. The performance of the MCAE model will be evaluated with respect to how much the feature maps vary across the domains, where a low variance is desirable. For these experiments, the synthetic dataset is divided into train and test sets following an 80\%-20\% split, resulting in approximately 16,000 and 4,000 image-patch triplets for each set respectively. In this section, domain A refers to the TCGA-COAD dataset, domain B refers to the synthetic NCT-CRC dataset and domain C refers to the synthetic NTSPH dataset.

\subsubsection{Normalised Feature Similarity}
\label{sec:normalised_feature_similarity}
The primary method of comparison used in this section is feature similarity. For example, given the feature maps $Z_S$ from a source domain and the feature maps $Z_T$ from a target domain, the MCAE system will be deemed successful if the feature vectors from both domains are similar whilst being rich in information about the underlying tissue. One approach is to measure the mean-squared error between the two feature vectors. The standard mean-squared error provides a useful measure of feature similarity but it can be misleading when comparing two different models because it does not account for the scale of the two different feature spaces. For example, the MCAE model could produce a feature space with a standard deviation of 0.1, whereas the comparative model StaNoSA could produce a feature space with a deviation of 1.0. This would make the comparison unfair as the relative differences in representations in the StaNoSA model would have a much higher absolute contribution. For this reason, when making comparisons on the feature spaces, each dimension of each feature space is normalised to have zero-mean and unit variance. For example, a normalised feature map $\hat{Z} \in \mathbb{R}^{h \times w \times n}$ where $h$ is the height, $w$ is the width and $n$ in the channel dimensionality of the feature map, can be computed from some feature map $Z$ as per Equation \ref{eq:normalised_feature_map}
\begin{equation}
    \hat{Z} = \frac{Z - \mu}{\sigma + \epsilon}
    \label{eq:normalised_feature_map}
\end{equation}
where $\mu \in \mathbb{R}^{n}$ is the mean vector representing the average value of each feature dimension in $Z$ and $\sigma \in \mathbb{R}^{n}$ is the channel-wise standard deviation vector, representing the standard deviation of each channel in $Z$. The normalised feature mean-squared-error (NFMSE) between two normalised feature maps $\hat{Z}_A$ and $\hat{Z}_B$ can then be calculated as per Equation \ref{eq:normalised_feature_mse}
\begin{equation}
    \text{NFMSE}(\hat{Z}_A, \hat{Z}_B) = \frac{1}{h \times w \times n} \sum_{i}^{h} \sum_{j}^{w} \sum_{k}^{n} \left( \hat{Z}_A^{i,j,k} - \hat{Z}_{B}^{i,j,k} \right)^2
    \label{eq:normalised_feature_mse}
\end{equation}

\subsubsection{Experimental Setup}
Both the MCAE and StaNoSA models are trained for 300 epochs using the Adam optimiser \citep{kingma2014adam} using a learning rate of 0.0002. For the MCAE model, three auto-encoders are defined and each is trained in tandem using image triplets from the test subset of the synthetic dataset described in Section \ref{sec:cyclegan_generated_synthetic_dataset}. This results in three trained auto-encoders, one for each domain in the synthetic dataset. For the StaNoSA model, each domain in the synthetic dataset is preprocessed according to the steps outlined in \citep{janowczyk2017stain} that includes global contrast normalisation and ZCA whitening of each 8$\times$8 RGB patch. The StaNoSA model is trained on the training subset of the synthetic dataset using the TCGA-COAD domain only. This is in line with the training instructions outlined in \citep{janowczyk2017stain} where the auto-encoder is trained on a single domain.

\subsubsection{Feature Similarity Results}
\begin{figure}
    \centering
    \begin{subfigure}[t]{0.45\linewidth}
        \includegraphics[width=0.99\linewidth]{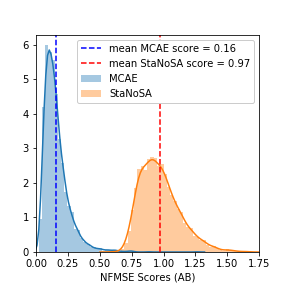}
        \caption{ \ }
        \label{fig:NFMSE_DISTRIBUTIONS_AB}
    \end{subfigure}
    \begin{subfigure}[t]{0.45\linewidth}
        \includegraphics[width=0.99\linewidth]{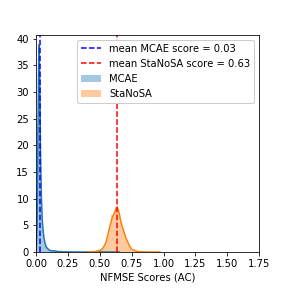}
        \caption{ \ }
        \label{fig:NFMSE_DISTRIBUTIONS_AC}
    \end{subfigure}
    \begin{subfigure}[t]{0.45\linewidth}
        \includegraphics[width=0.99\linewidth]{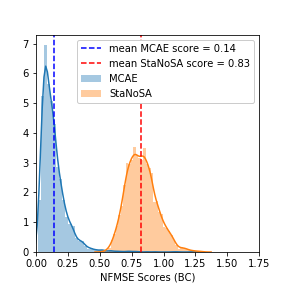}
        \caption{ \ }
        \label{fig:NFMSE_DISTRIBUTIONS_BC}
    \end{subfigure}
    \caption{Normalised Features Mean-Squared Error (NFMSE) score distributions between feature vectors in domains A and B (a), A and C (b) and B and C (c).}
     \label{fig:NFMSE_Distributions}
\end{figure}

Once the training process is complete, the triplets contained in the testing set are mapped to the feature space using the appropriate encoding function for each domain. For each triplet, the NFMSE is then calculated between each pair-configuration (A and B, A and C, B and C) by squaring the differences between feature vectors for corresponding pixels and then dividing by the number of pixels. The distribution of these NFMSE scores is shown in Figure \ref{fig:NFMSE_Distributions} which shows that the MCAE model produces feature vectors which are much more similar across domains than the StaNoSA model. When comparing features from domain A with those from domain B (Figure \ref{fig:NFMSE_DISTRIBUTIONS_AB}), the MCAE model achieved an average NFMSE score of 0.15819 compared to 0.97128 for StaNoSA. Comparing domains A and C (Figure \ref{fig:NFMSE_DISTRIBUTIONS_AC}) yields 0.03257 and 0.63196 for MCAE and StaNoSA respectively. Finally, comparing domains B and C (Figure \ref{fig:NFMSE_DISTRIBUTIONS_BC}) gives 0.13757 for MCAE and 0.82684 for StaNoSA. These results demonstrate a significant improvement in feature similarity when using the MCAE model over StaNoSA. It is likely that this is due to the active-feature normalisation approach taken in the MCAE model where each auto-encoder is actively learning feature representations that are domain-invariant whereas the StaNoSA model relies on global contrast normalisation and ZCA whitening to reduce the impact of the inter-domain variances.

\subsection{Tissue Classification}
\label{sec:tissue_classification}
In Section \ref{sec:domain_invariance}, it was shown that the MCAE model produces feature representations that are statistically more similar across domains than the StaNoSA model. In this section, the usefulness of the feature representations are evaluated with respect to a tissue classification task. This evaluation will be done on the NCT-CRC dataset discussed in Section \ref{sec:data_domains}.

Each image in the NCT-CRC dataset is assigned one of nine class labels: adipose (ADI), background (BACK), debris (DEB), lymphocytes (LYM), mucus (MUC), smooth muscle (MUS), normal colon mucosa (NORM), cancer-associated stroma (STR) and colorectal adenocarcinoma epithelium (TUM). The dataset contains 100,000 images and is divided by a 75\%-5\%-20\% training-validation-testing split, resulting in 74997, 4997 and 20006 images in the training, validation and testing subsets respectively. 

The efficacy of the MCAE and StaNoSA models will be determined by mapping images from the NCT-CRC dataset into feature maps using the encoder portion of the MCAE or StaNoSA models as feature extractors. The feature extractor models will be frozen and identically structured classifiers will be trained on top of them using the training subset. Given that the feature extraction models are frozen for this experiment, their performance on this task can be used to estimate how well each model generalises to new data. 

\subsubsection{Training Process}
Pre-trained MCAE and StaNoSA models are taken from the experiments performed in Section \ref{sec:normalised_feature_similarity}. The StaNoSA model has been trained on data from the TCGA domain. For the MCAE model, there are three auto-encoders to choose from; the auto-encoder that was trained on data from the TCGA domain is chosen to match the StaNoSA model as much as possible. For both models, the decoder portion is discarded and the encoder portion is selected for use as a feature extractor. A classification module is comprised of two convolution layers, each followed by the leaky-relu activation function and an adaptive pooling layer at the end which reduces the spatial dimensions of the feature map such that the output can be easily represented as a class-wise vector of logits. For the rest of this section, the terms MCAE and StaNoSA will refer to the encoder portions of each model (and any model-specific pre-processing), plus the classification module.

The models are trained in tandem for 100 epochs. Each image is divided up into non-overlapping patches of 8$\times$8 RGB pixels, such that each 224x224 pixel input image is divided up into 28x28 patches. Each of these patches is reshaped into a single 192-element vector and pre-processed according to the requirements of each model. For example, before passing the vectors to the MCAE model, the vector is simply normalised to the range [-1,1], whereas for the StaNoSA model the vectors are linearly transformed using a pre-calculated ZCA whitening matrix. Once the vectors have been separately pre-processed, they are processed by the MCAE and StaNoSA models to produce class-wise logit vectors. These logit vectors are used to calculate the cross-entropy loss and the weights of the classification modules are updated by back-propagation and the Adam optimiser \citep{kingma2014adam} with a learning rate of 0.0002. At any given training iteration, the MCAE and StaNoSA models receive exactly the same input (with the exception of the model-specific pre-processing), however it is important to note that they are still completely separate.

\subsubsection{Results}
Once trained, the unseen test subset of the NCT-CRC dataset is processed by each model. The class-wise precision, recall and f1-scores for the MCAE and StaNoSA models can be seen in Tables \ref{tab:MCAE_tissue_clf_results} and \ref{tab:StaNoSA_tissue_clf_results} respectively. The f1 scores in bold highlight the best performance in each class across both tables.

It can be seen that the MCAE model presents a 5 percentage-point improvement over the StaNoSA model across all classes. However, NCT-CRC dataset is not balanced and therefore the `weighted avg` of each metric is important as it takes into account the proportions of each class. In this case, the MCAE is still able to achieve a 5 percentage-point improvement over the StaNoSA model when examining the class-weighted average f1-score. Overall, these results suggest that the MCAE model generalises better to new data compared to the StaNoSA model. The results of this experiment suggest that the features produced by the MCAE model are indeed useful and able to extract useful information about the tissue structures contained in histopathology images.

\begin{table}[ht]
\centering
\caption{MCAE tissue classification results on the unseen, NCT-CRC test subset.}
\label{tab:MCAE_tissue_clf_results}
\vspace{0.5em}
\begin{tabular}{lllll}
\hline
class        & precision & recall & f1-score & support  \\ 
\hline
ADI          & 0.77      & 0.61   & \textbf{0.68}     & 1753     \\
BACK         & 0.81      & 0.80   & 0.80     & 1780     \\
DEB          & 0.89      & 0.88   & 0.89     & 2082     \\
LYM          & 0.91      & 0.96   & 0.93     & 2114     \\
MUC          & 0.63      & 0.76   & \textbf{0.69}     & 2303     \\
MUS          & 0.66      & 0.58   & \textbf{0.62}     & 2090     \\
NORM         & 0.74      & 0.86   & \textbf{0.80}     & 2864     \\
STR          & 0.97      & 0.97   & \textbf{0.97}     & 2312     \\
TUM          & 0.89      & 0.74   & \textbf{0.81}     & 2708     \\ 
\hline
accuracy     &           &        & \textbf{0.80}     & 20006    \\
weighted avg & 0.81      & 0.80   & \textbf{0.80}     & 20006    \\ 
\hline
             &           &        &          &         
\end{tabular}
%
\centering
\caption{StaNoSA tissue classification results on the unseen, NCT-CRC test subset.}
\label{tab:StaNoSA_tissue_clf_results}
\vspace{0.5em}
\begin{tabular}{lllll}
\hline
class        & precision & recall & f1-score & support  \\ 
\hline
ADI          & 0.69      & 0.35   & 0.47     & 1753     \\
BACK         & 0.75      & 0.88   & \textbf{0.81}     & 1780     \\
DEB          & 0.93      & 0.92   & \textbf{0.93}     & 2082     \\
LYM          & 0.96      & 0.96   & \textbf{0.96}     & 2114     \\
MUC          & 0.55      & 0.76   & 0.64     & 2303     \\
MUS          & 0.59      & 0.36   & 0.45     & 2090     \\
NORM         & 0.64      & 0.84   & 0.73     & 2864     \\
STR          & 0.98      & 0.87   & 0.92     & 2312     \\
TUM          & 0.79      & 0.75   & 0.77     & 2708     \\ 
\hline
accuracy     &           &        & 0.75     & 20006    \\
weighted avg & 0.76      & 0.75   & 0.75     & 20006    \\
\hline
\end{tabular}
\end{table}

\section{Conclusion}\label{sec:conclusion}
This work presented the Multi-Channel Auto-Encoder which is a novel extension of the Dual-Channel Auto-Encoder model that makes use of more than two data domains. Additionally, this work demonstrated that the active feature normalisation approach used by the MCAE model is superior to other statistical normalisation methods like ZCA whitening.

A limitation of the MCAE and DCAE models is that they require aligned data from multiple domains in order to learn normalised feature representations. In order to alleviate the impact of this limitation, a novel synthetic dataset was generated using a self-attentive CycleGAN based on the work outlined in \citep{shrivastava2019self}. This dataset was generated by fitting 3 CycleGAN models to 3 domains of unaligned tissue image data. The trained CycleGANs were then used to produce a novel dataset of aligned tissue image triplets where each triplet represents the same tissue patch but with the staining appearance of other domains.

In Section \ref{sec:synthetic_dataset_evaluation}, the synthetic dataset was evaluated with respect to the quality of appearance normalisation and suitability for use in further experiments. It was shown in Section \ref{sec:synthetic_dataset_evaluation} (Figures  \ref{fig:cxcy_distributions}, \ref{fig:density_comparison} and Table \ref{tab:density_similarity}) that the CycleGAN models are able to capture and transfer the colour appearance of different tissue domains.

In Section \ref{sec:domain_invariance}, the abilities of the MCAE and StaNoSA models to produce domain-invariant feature representations were evaluated on the synthetic dataset. Using the Normalised Feature Mean-Squared Error (NFMSE) metric, the MCAE model is able to produce features that vary much less across domains than the StaNoSA model. This is likely due to the active feature normalisation approach in the MCAE model that uses feature-based loss functions to actively encourage the learning of normalised feature representations of images as opposed to the StaNoSA model, which makes use of preprocessing techniques such as global contrast normalisation and ZCA whitening in order to reduce the impact of inter-domain variances.

In Section \ref{sec:tissue_classification}, the ability of the MCAE and StaNoSA models to generalise to a new dataset and new task was evaluated using the NCT-CRC tissue classification dataset. The MCAE and StaNoSA models were used as pre-trained feature extractors, the outputs of which were fed into separate classification modules. The results showed that the classifier that used MCAE as a feature extractor outperformed the model using StaNoSA as a feature extractor by 5 percentage-points with respect to f1-score.

These results suggest that the active-feature normalisation approach used in the MCAE model results in features that generalise better to novel tasks. Despite promising results, the MCAE model is limited by the lack of convolutional layers. Explicitly extracting patches and processing them with fully connected layers is slower than using standard convolutions and so the exact architecture used in this paper is not recommended. Instead, the concept of active feature normalisation should be applied to convolutional feature extractors. 

Future work will involve adapting the MCAE model to utilise convolutional layers for greater efficiency and to develop a novel stain augmentation method that can be used to simulate various staining and scanning conditions on-the-fly, thus reducing the need to use CycleGANs to synthesise the data beforehand.

\section*{Acknowledgments}
This work was financially supported by InvestNI, the Department for Employment and Learning (Northern Ireland) and the Natural Science Foundation of Jiangsu Province, China (Grant no. BK20170443).










\end{document}